\documentclass[reprint,amsmath,amssymb,aps,prb,superscriptaddress,longbibliography]{revtex4-1}

\pdfoutput=1

\usepackage{hyperref,url}
\usepackage{amsmath}
\usepackage{amsfonts}
\usepackage{mathtools}
\usepackage[capitalise]{cleveref}
\usepackage{placeins}
\usepackage[version=4]{mhchem} 
\hypersetup{breaklinks,colorlinks=true,linkcolor=blue,citecolor=blue,urlcolor=blue,filecolor=blue}
\usepackage{graphicx}
\usepackage{dcolumn}
\usepackage{rotating}
\usepackage{etoolbox}

\newcommand\mat\mathbf

\newcommand{\Columbia}{\affiliation{Department of Chemistry, Columbia University, New York, NY, USA}}
\newcommand{\Google}{\affiliation{Google Quantum AI, Venice, CA, USA}}
\newcommand{\Cal}{\affiliation{Berkeley Quantum Information \& Computation Center, University of California, Berkeley, CA, USA}}

\begin{document}

 \author{Joonho Lee}
 \email{jl5653@columbia.edu}
 \Columbia
 \Google
 \author{David R.~Reichman}
 \Columbia
 \author{Ryan Babbush}
 \Google
 \author{Nicholas C.~Rubin}
 \Google
 \author{Fionn D.~Malone}
 \Google
 \author{Bryan O'Gorman}
 \Cal\thanks{Present address: IBM Quantum, IBM T.J. Watson Research Center, Yorktown Heights, NY, USA}
 \author{William J.~Huggins}
 \Google

\title{Response to ``Exponential challenges in unbiasing quantum Monte Carlo algorithms with quantum computers''}

\begin{abstract}
A recent preprint  by Mazzola and Carleo\cite{Mazzola1013Oct}  numerically investigates exponential challenges that can arise for the QC-QMC algorithm introduced in our work, ``Unbiasing fermionic quantum Monte Carlo with a quantum computer.''\cite{Huggins2022Mar} 
As discussed in our original paper, we agree with this general concern.
However, here we provide further details and numerics to emphasize that the prospects for practical quantum advantage in QC-QMC remain open.
The exponential challenges in QC-QMC are dependent on (1) the choice of QMC methods, (2) the underlying system, and (3) the form of trial and walker wavefunctions.
While one can find difficult examples with a specific method, a specific system, and a specific walker/trial form, for some combinations of these choices the approach is potentially more scalable than other near-term quantum algorithms. Future research should aim to identify examples for which QC-QMC enables practical quantum advantage.
\end{abstract}
\maketitle
\newpage

{\it Introduction.} QC-QMC relies on estimating wavefunction amplitudes 
$\langle \Psi_T | \phi \rangle$ on quantum computers where $|\Psi_T\rangle$ is a trial wavefunction and
$|\phi\rangle$ is a walker (statistical sample) wavefunction.
Quantum computers can only efficiently approximate these amplitudes up to an additive error $\epsilon$, meaning $\langle \Psi_T | \phi \rangle + \mathcal O(\epsilon)$, while $\langle \Psi_T | \phi \rangle$ decays to zero exponentially with system size. 
Given these facts, the number of measurements necessary to keep
the signal-to-noise ratio fixed will grow exponentially with system size.
Nonetheless, there is no evidence that suggests that this exponential challenge is always the computational bottleneck in QC-QMC when considering finite size problems.
Furthermore, the central motivation of QC-QMC is to use $|\Psi_T\rangle$ that poses exponential challenges to known classical algorithms such that even estimating $\langle \Psi_T | \phi \rangle$ up to an additive error $\epsilon$ scales exponentially with system size. Interested readers are referred to Supplementary Section F of our original work for more details about scaling.\cite{Huggins2022Mar}

With these considerations in mind, 
we consider the specific example 
investigated by Mazzola and Carleo,
the one-dimensional (1D) transverse field Ising model (TFIM) with $L$ spins under periodic boundary conditions,
\begin{equation}
\mathcal H
=
-J \sum_{k=1}^L \sigma^z_{k} \sigma^z_{k+1} - \Gamma\sum_{k=1}^L\sigma_k^x,
\end{equation}
where $J$ and $\Gamma$ are the Hamiltonian parameters and $\sigma^z_k$ and $\sigma^x_k$ are the Pauli $Z$- and $X$-operators, respectively, for the $k$-th spin.

{\it Method dependence}
We presented QC-QMC as a general framework that combines wavefunction amplitudes estimation on the quantum computer with constrained QMC simulations on the classical computer. For a given system, the performance of QC-QMC can only be analyzed after making a specific choice of QMC method. In our work,\cite{Huggins2022Mar} we considered auxiliary-field quantum Monte Carlo (AFQMC)\cite{Zhang2003Apr} whereas Ref. \citenum{Mazzola1013Oct} used Green's functiom Monte Carlo (GFMC).\cite{Becca2017Nov} Hence, these are two separate methods, QC-AFQMC and QC-GFMC.
If one were to apply QC-AFQMC to the 1D TFIM, one would map the model to a fermionic system.\cite{Pfeuty1970Mar} The resulting fermionic Hamiltonian is non-interacting, for which QC-AFQMC (or AFQMC with any trial wavefunction) is exact without QMC sampling.
This highlights how the numerical examples presented in Ref. \citenum{Mazzola1013Oct} are only relevant to QC-GFMC, not necessarily to other flavors of QC-QMC.

{\it System dependence.}
The systems that we studied in our work\cite{Huggins2022Mar} consist of electronic degrees of freedom in chemical systems, \ce{H4}, \ce{N2}, and diamond. We tuned the bond distance between atoms to control the difficulty of the problems similarly to tuning $\Gamma/J$ in the 1D TFIM.
For instance, the 1D TFIM with $\Gamma/J=1$ is the limit where the ground state has overlaps with many product states in the $Z$-basis, rendering apparent difficulties despite its non-interacting nature. On the other hand, the $\Gamma/J=0$ limit is trivial for QC-GFMC when starting from $|\uparrow\uparrow\cdots\uparrow\rangle$ or $|\downarrow\downarrow\cdots\downarrow\rangle$, the ground states in this limit. This suggests that the exponential challenges in QC-GFMC for the 1D TFIM must depend on the system (i.e., $\Gamma/J$).
To explore this, we consider an example of $\Gamma/J = 0.5$ with a trial wavefunction,
\begin{equation}
|\Psi_T\rangle=\exp\left[\lambda \sum_i \sigma^x_i\right]\left(|\uparrow\uparrow\cdots\uparrow\rangle+|\downarrow\downarrow\cdots\downarrow\rangle\right),
\end{equation}
with $\lambda = 0.127$. This trial wavefunction reproduces the exact ground state energy better than 99.8\%~for $L$ from 6 to 12.

\begin{figure*}
    \centering
    \includegraphics[scale=0.75]{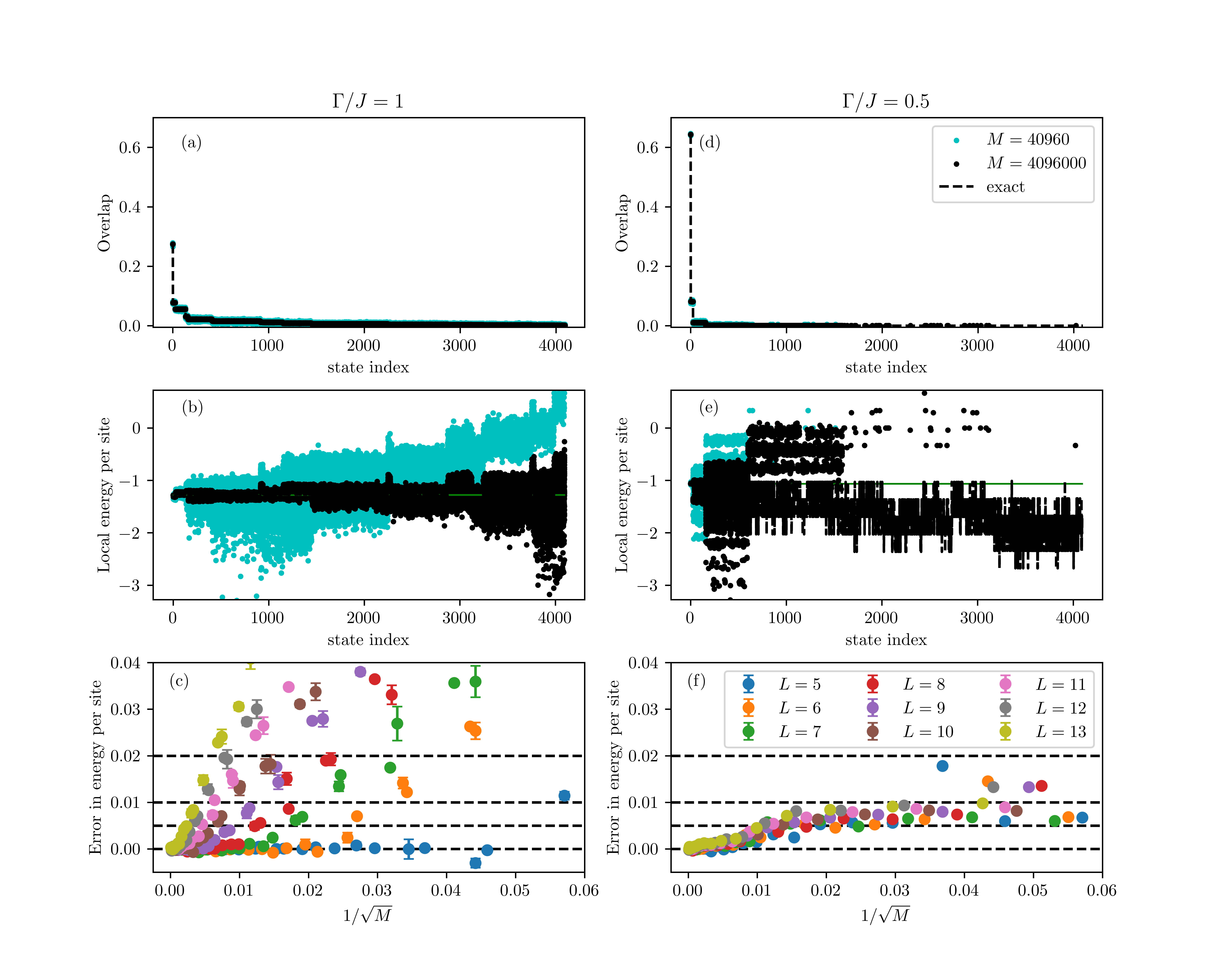}
    \caption{(a) Overlaps for each of the computational basis states and (b) local energy per site for the system size $L=12$, $\Gamma/J=1.0$ 1D TFIM for different values of numbers of measurements ($M$). (c) Error in GFMC energy per site as a function of $M^{-1/2}$ for different system sizes for the $\Gamma/J=1.0$ 1D TFIM. 
    (d), (e), and (f) are the same
    plots for $\Gamma/J=0.5$.
    For overlap and local energy, 16 independent simulations were run with $M$ samples each time, sampling computational basis states exactly from $|\Psi_T|^2$.
    GFMC results were averaged over 128 independent runs. Black dotted lines in (c) and (f) are drawn to guide the eye, corresponding to 0.02, 0.01, 0.005, and 0, respectively.
    Comparing (c) an (f), we observe that system size ($L$) dependence as a function of the number of measurements varies significantly depending on $\Gamma/J$.
    }
    \label{fig:h05}
\end{figure*}

Following Ref. \citenum{Mazzola1013Oct}, we provide similar numerics on the $\Gamma/J=0.5$ 1D TFIM, the results of which are presented in \cref{fig:h05} (d)-(f) along with (a)-(c) for $\Gamma/J=1.0$ studied in Ref. \citenum{Mazzola1013Oct} for comparison.
The overlap distribution is sharper for $\Gamma/J=0.5$ (\cref{fig:h05} (d)) than $\Gamma/J=1.0$ (\cref{fig:h05} (a)), yielding more accurate local energies for the states with a higher overlap with $|\Psi_T\rangle$.
For $\Gamma/J=0.5$ (\cref{fig:h05} (f)), we find weak system size dependence of the GFMC total energy  with the number of samples, showing stark differences from that of $\Gamma/J=1.0$ (\cref{fig:h05} (c)).
We can conclude therefore that in this regime QC-QMC may be a useful technique, and we cannot generally preclude its efficiency based on the $\Gamma/J=1$ TFIM example.

{\it Walker and trial wavefunctions dependence.}
Since our central quantum task is to compute the overlap between trial and walker wavefunctions, the exponential challenge in this task is inherently dependent on the form of trial and walker wavefunctions.
In the limit of $|\phi\rangle = |\Psi_T\rangle = |\Psi_0\rangle$ where $|\Psi_0\rangle$ is the exact ground state, the resulting QC-QMC energy would be exact within an additive error $\epsilon$ and no QMC sampling is required. While we do not have the zero variance principle,\cite{Becca2017Nov} our measurement overhead to compute the local energy would not be exponential scaling in this limit.
More generally, if we use the initial walker wavefunction to be the same as a sophisticated trial wavefunction $|\Psi_T\rangle$, the resulting overlap distribution will become much more favorable for QC-QMC compared to that of product state walkers. 

\begin{figure*}  
    \centering
    \includegraphics[scale=0.85]{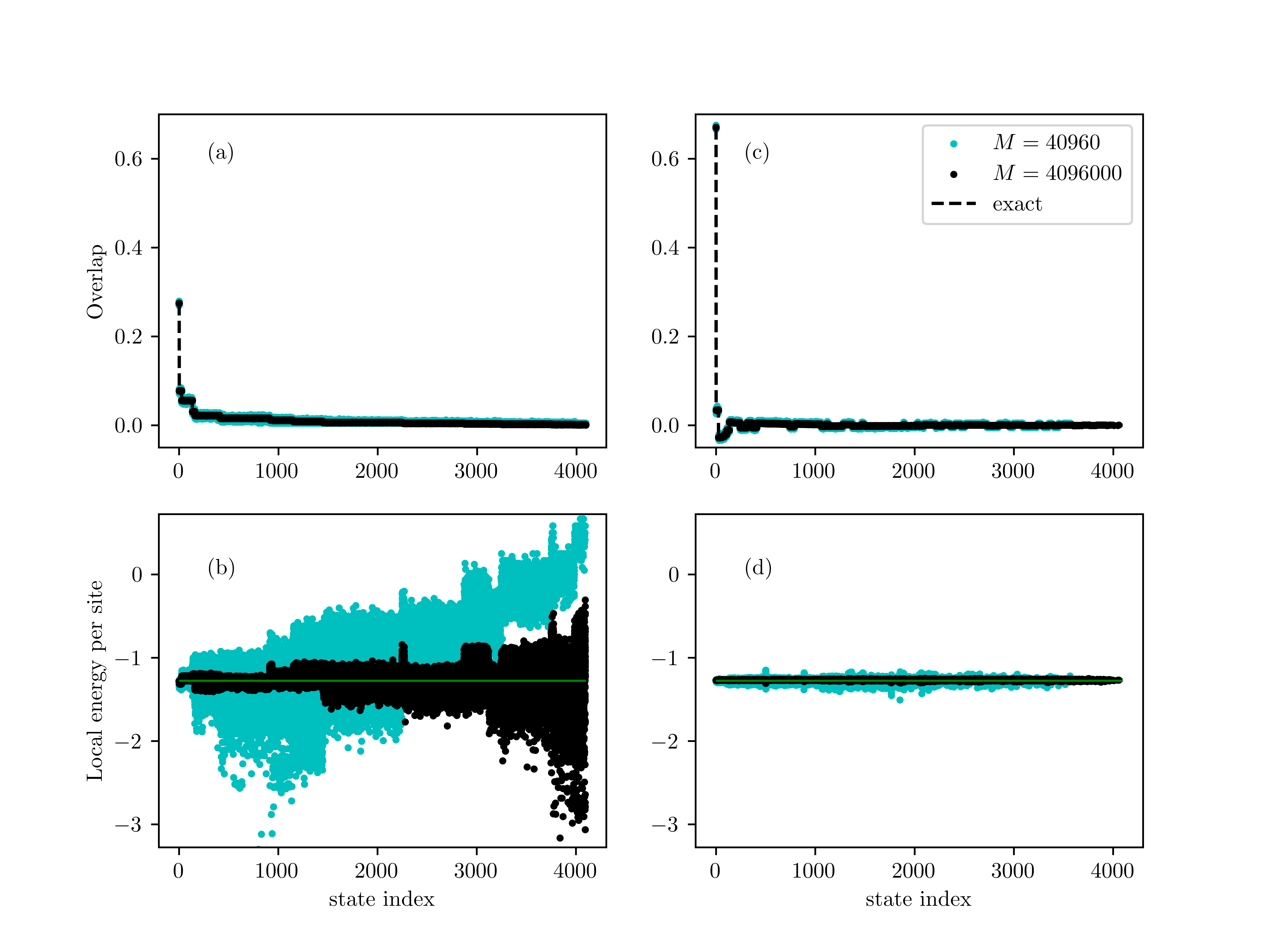}
    \caption{Comparison between overlaps and local energy per site  for the $L=12$, $\Gamma/J=1$ 1D TFIM with different types of walker wavefunctions for different values of number of measurements ($M$). For (a) and (b) we used standard (single computational basis states) GFMC walkers while for (c) and (d) more sophisticated walker wavefunctions, generated by spin flips from $|\Psi_{MC}\rangle$, were used.
    For (a) and (b), we used $|\Psi_T\rangle = |\Psi_{MC}\rangle$ 
    and
    for (c) and (d) we employed $|\Psi_T\rangle = e^{-0.05 \mathcal H} |\Psi_{MC}\rangle$.
    For overlap and local energy, 16 independent simulations were run with $M$ samples each time, sampling computational basis states exactly from $|\Psi_T|^2$. For (c), only the magnitude was obtained from sampling, not the sign.
    }
    \label{fig:fancy_walkers}
\end{figure*}

To be concrete, we take the trial wavefunction used in Ref. \citenum{Mazzola1013Oct} (denoted by $|\Psi_{MC}\rangle$) as the initial wavefunction and perform a similar overlap and local energy analysis on $\Gamma/J = 1$. 
Our walkers are non-orthogonal states generated by 1, 2, $\ldots$, and $L$ spin flips from $|\Psi_{MC}\rangle$.
While this type of QMC is often classically exponential-scaling, this could suppress the exponential measurement overhead in QC-QMC greatly.
To investigate this idea, we plot in \cref{fig:fancy_walkers} (a) and (b) the overlap and local energy using the product state walkers and contrast this to our more complex walkers, the results of which are shown in panels (c) and (d). 
It is clear from \cref{fig:fancy_walkers} that the use of sophisticated walkers serves to sharpen the distribution of weight around a handful of states and also greatly reduces the variance in the local energy per site.
Given these numerics, we can conclude that QC-GFMC of this type (and QC-QMC more generally) offers the possibility of postponing the exponential-scaling measurement bottleneck that can quickly affect the simpler version.

The use of sophisticated walkers complicates the implementation of QC-GFMC (or QC-QMC more generally) and may introduce a sign problem in otherwise sign-problem free models, as indicated in \cref{fig:fancy_walkers}(b).
Despite these tradeoffs, approaches of this kind may prove useful in increasing the scope of QC-QMC.


{\it Comparison to other quantum algorithms.} 
Due to the QMA-hardness, all known ground state algorithms work accurately and efficiently for some systems but not for others. 
Without actually running these algorithms for specific systems on the quantum computer, it is difficult to make a precise cost comparison between different quantum algorithms, such as QC-QMC, variational quantum eigensolver (VQE),~\cite{Peruzzo2014Jul}, quantum phase estimation (QPE),~\cite{Aspuru-Guzik2005Sep} and quantum imaginary time-evolution (QITE).~\cite{Motta2020Feb} Like QC-QMC, all of these approaches have (different) regimes where we expect exponential scaling. In Ref. \citenum{Huggins2022Mar}, we presented QC-QMC as a more noise-resilient and efficient alternative to these approaches, especially for a near-term processor. Noise-resilience was evidenced by the actual accurate quantum simulation that we performed up to 16 qubits. The efficiency of QC-QMC depends on how small the individual amplitudes are and for our own quantum calculations these exponential challenges were not the bottleneck. 

As suggested in Ref. \citenum{Mazzola1013Oct}, QC-QMC may be fundamentally limited to targeting ground states that can be approximately sparsely represented in the basis of walker wavefunction. 
Even granting this assumption, our algorithm is more efficient in the quantum resources and measurement overhead than QPE and VQE when adding more basis functions while keeping the number of particles unchanged (see Supplementary Section C3 in Ref. \citenum{Huggins2022Mar}.) 
This is because QC-QMC obtains electron correlation outside the qubit space without any additional quantum resources or measurement overheads. In this sense, our algorithm offers a compelling advantage in scaling to the complete basis set limit when compared with  VQE, QPE, QITE, and other quantum algorithms.

{\it Conclusions.} We hope that our response unraveled some of the subtle differences between our original work\cite{Huggins2022Mar} and numerical results presented in Ref. \citenum{Mazzola1013Oct}.
We showed, by numerical simulations, that the exponential challenges raised by our work and numerically studied in Ref. \citenum{Mazzola1013Oct} inherently depend on the choice of (1) methods, (2) systems, and (3) trial and walker wavefunctions. 
We expect that identifying a specific combination of these details to demonstrate a practical quantum advantage in QC-QMC will remain an active research area in the future.

{\it Acknowledgement}
JL thanks Garnet Chan for helpful discussion and encouragement. 
We acknowledge computing resources from Columbia University's Shared Research Computing Facility project, which is supported by NIH Research Facility Improvement Grant 1G20RR030893-01, and associated funds from the New York State Empire State Development, Division of Science Technology and Innovation (NYSTAR) Contract C090171, both awarded April 15, 2010.
\bibliography{refs}

\end{document}